\documentclass[11pt]{article}
\usepackage{graphicx}
\usepackage{caption}
\usepackage{amsfonts}
\usepackage{amsmath}
\usepackage{amssymb}
\usepackage{flafter}
\usepackage[margin=0.7in]{geometry}

\begin{document}

\title{Quantitative and Qualitative Seismic Imaging and Seismic Inversion}

\author{August Lau and Chuan Yin}

\maketitle

\begin{abstract}
We consider seismic imaging to include seismic inversion.  Imaging could use approximate operator or time instead of depth. Processing in time is an important part of seismic imaging as well as processing in depth.
We can classify seismic imaging as quantitative versus qualitative methods.  Quantitative method uses numerical methods to find the solution whose modeled seismic data approximates the input seismic records.  Then we will progress to qualitative methods which have three aspects.  The first aspect will be topology and geometry.  The second aspect is semigroup method. The third aspect is to use non-differentiable solution.
\end{abstract}

\subsection*{QUANTITATIVE METHOD}

We will first discuss seismic imaging (inverse problem) as direct method and indirect method.  We can write the symbolic equation:
$$Ax = y,$$
where $A$ is the operator, $x$ is the solution, $y$ is the input data.
Direct inversion approximates the operator and indirect inversion approximates the data. 

Direct inversion is to find an operator $B$ so that
$$x = By,$$
where $A$ and $B$ can be viewed as dual operators to each other.

Indirect inversion is to find $x$ so as to minimize $$L_{p} (Ax-y).$$
For example, if $p$ is 2, then $L_{2}$ is the least square norm attributed to Gauss.

\subsection*{NULLSPACE}

Regardless of the inversion method,  nullspace exists in which a real solution combined with a nullspace vector could yield the same seismic data.  For real field data,  we could assume that the real data is the same as modeled data if they are within a small error.  This is the non-unique nature of the inverse problem.

\subsection*{Riemann-Lebesgue Lemma}

The Riemann-Lebesgue Lemma states that if the Lebesgue integral of $f$ is finite, then the Fourier transform of $f$ satisfies
$$ F(k):=\int_{R^d} f(x) exp(-ikx)dx \Rightarrow 0 $$
as $$ k \Rightarrow \infty . $$
This is equivalent to stating the following for 1-D real functions $f:[a,b] \rightarrow R$ 
$$ \lim_{k \rightarrow \infty} \int^b_a f(x) \sin (kx) dx = \lim_{k \rightarrow \infty} \int^b_a f(x) \cos (kx) dx = 0 $$

A practical implication of the Riemann-Lebesgue Lemma on geopysical inversion can be demonstrated with the familiar 1-D convolutional model, which can be written as the  following
$$ d(t) = w(t) * r(t) = \int^t_0 w(t-\tau) r(\tau) d\tau, $$ 
where $w(t)$ is the wavelet function, $r(t)$ the reflectivity function, and $d(t)$ the seismogram. Acoording to the Riemann-Lebesgue Lemma, a different reflectivity function, $r_0(t)$, 
$$ r_0 (t) = r(t) + \alpha \sin (\beta t) $$
would also yield the identical seismogram,
because 

\begin{eqnarray*}
\lim_{\beta \rightarrow \infty} \int^t_0 w(t-\tau) [ r(\tau) + \alpha \sin (\beta \tau) ] d\tau & = &  \\
                                                               \int^t_0 w(t-\tau) r(\tau) d\tau & = & d(t). 
\end{eqnarray*}

\begin{figure}
\centering
  \includegraphics[width=5in]{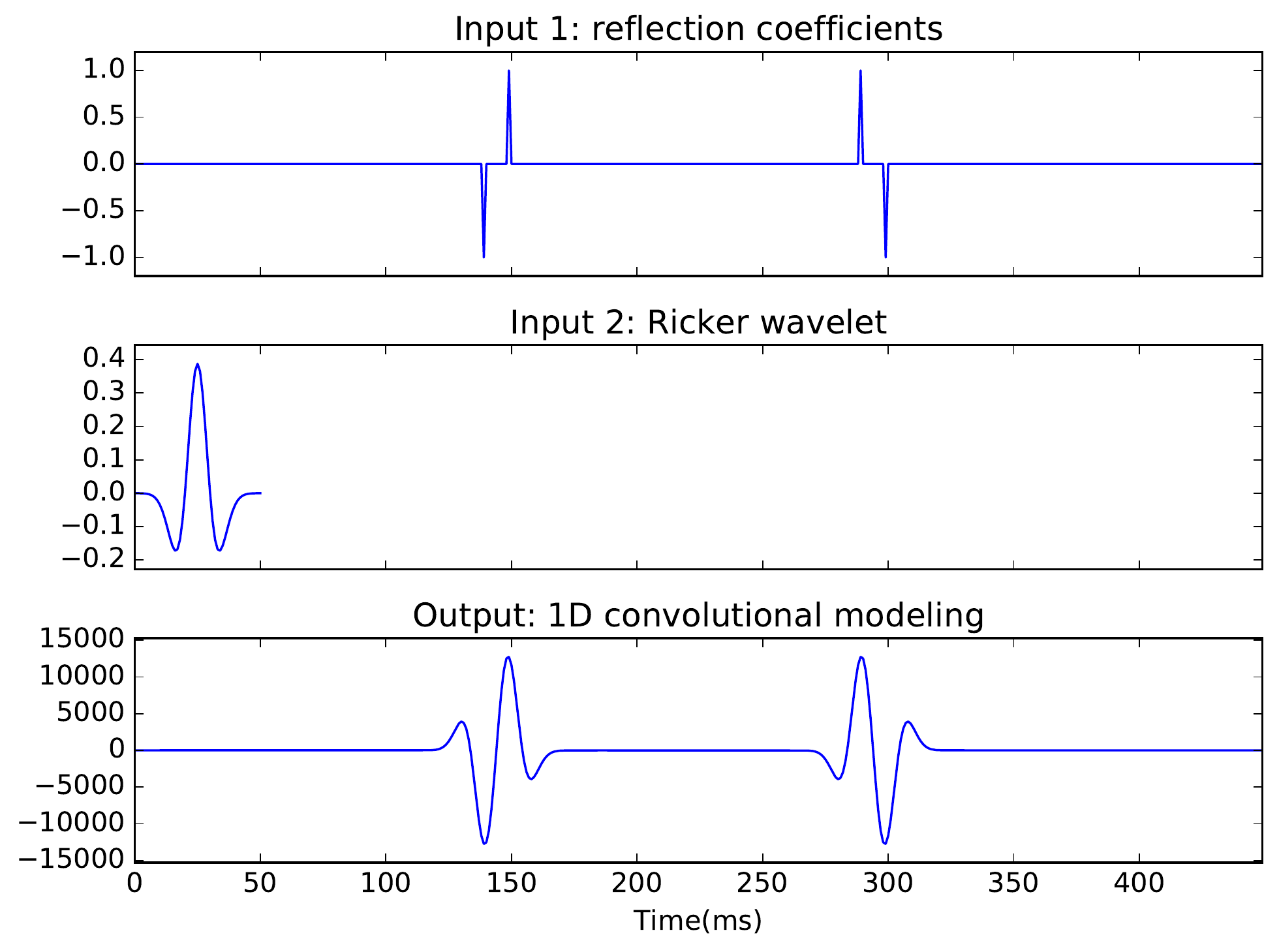}
\caption{Null space example - forward modeling using 1D convolutional modeling, $\alpha=0.0$,$\beta=0.0$}
\label{1dconv0}
\end{figure}

\begin{figure}
\centering
  \includegraphics[width=5in]{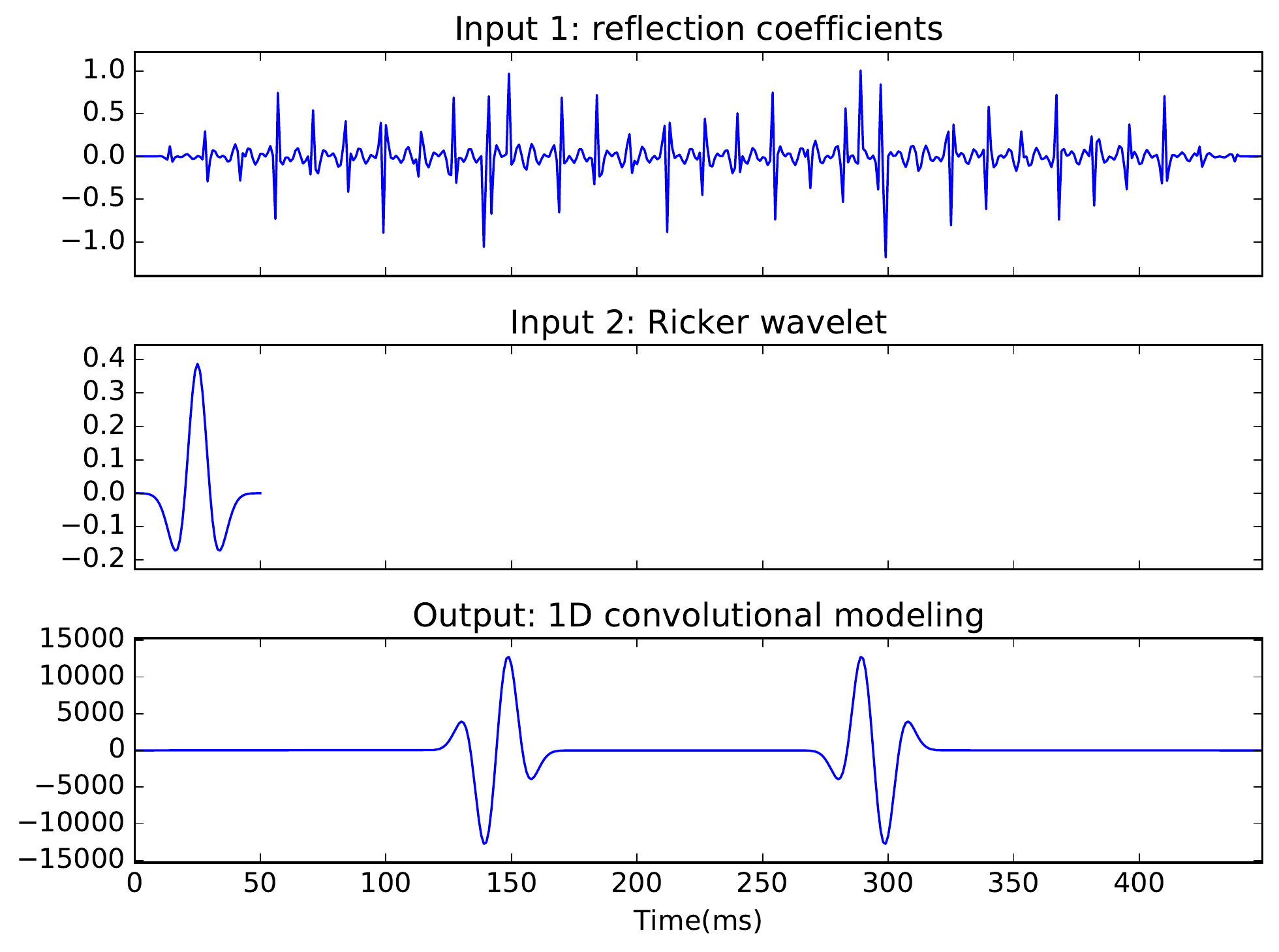}
\caption{Null space example - forward modeling using 1D convolutional modeling, $\alpha=0.1$,$\beta=1.0$ and $N=9$ in Eq.(1).}
\label{1dconv1}
\end{figure}

\begin{figure}
\centering
  \includegraphics[width=5in]{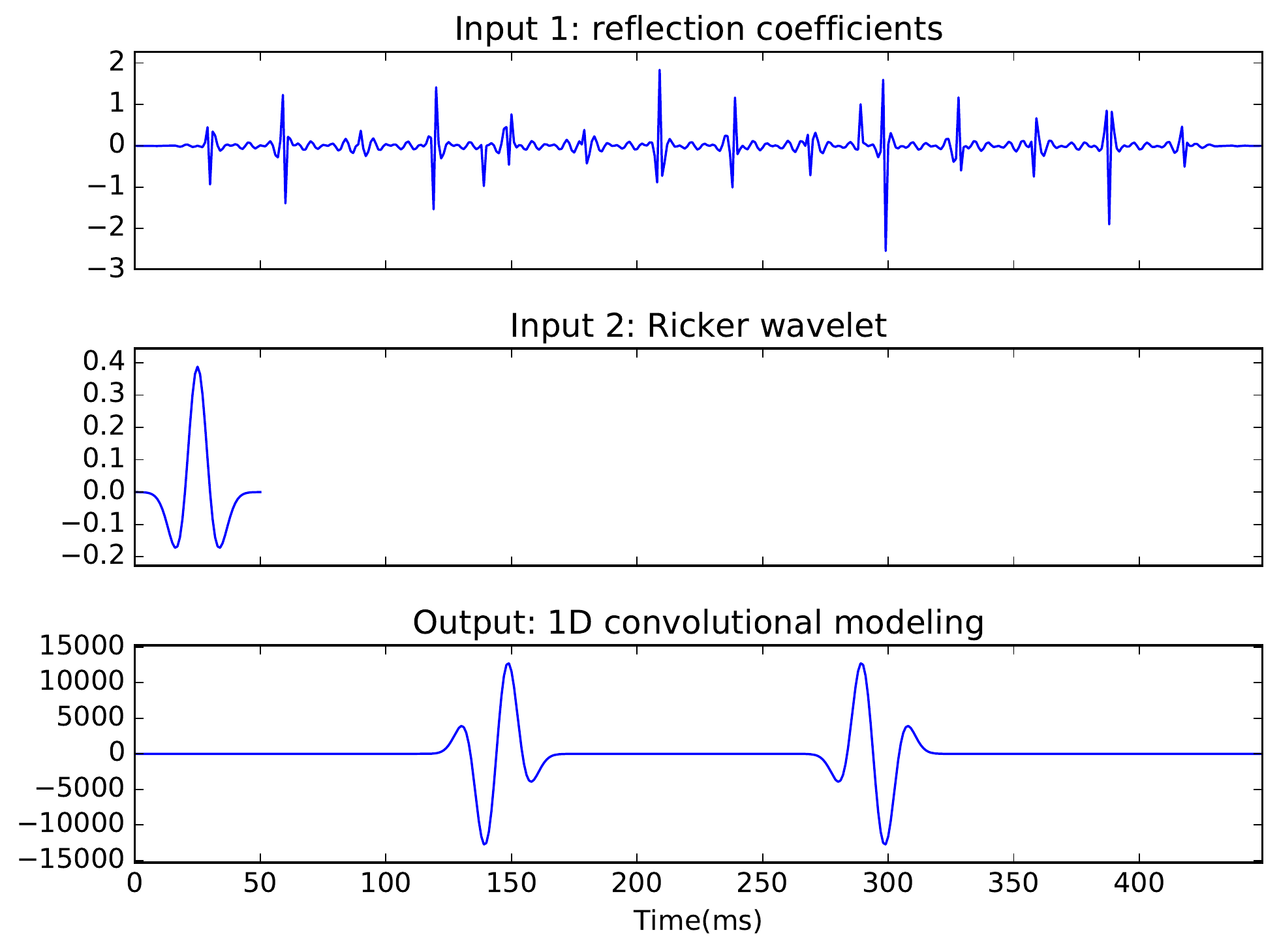}
\caption{Null space example - forward modeling using 1D convolutional modeling, $\alpha=0.1$,$\beta=1.0$ and $N=19$ in Eq.(1).}
\label{1dconv2}
\end{figure}

Figures \ref{1dconv0}, \ref{1dconv1} and \ref{1dconv2} are numerical and graphical demonstrations of this analytical result, where Figure 1 uses the original reflectivity. 
Figures 2 and 3 use non-zero values for $\alpha$ and $\beta$, $\alpha=0.1$,$\beta=1.0$, and with the following form
\begin{equation}
r_1 (t) = r(t) + \alpha \sum\limits_{n=0}^N \sin [(1+\frac{4n}{N})\beta t],
\end{equation}
where $N=9$ for Figure 2, and $N=19$ for Figure 3.
This example is a reminder that there is a huge null space even for just a 1-D inversion using the convolutional model.

\subsection*{Other Theoretical Concerns}

Inverse operators exist only under certain and sometimes restrictive assumptions in the sense of operators of Hilbert space or operators of topological space.  Locally invertible operators might not be globally invertible. See Lau (1979) local homeomorphism paper and (1980) transformation group. This is independent of which numerical method is implemented.  It is not easy to meet the assumptions for typical seismic recording where active sources and receivers are at the surface.  

\begin{figure}
\centering
  \includegraphics[width=6.5in]{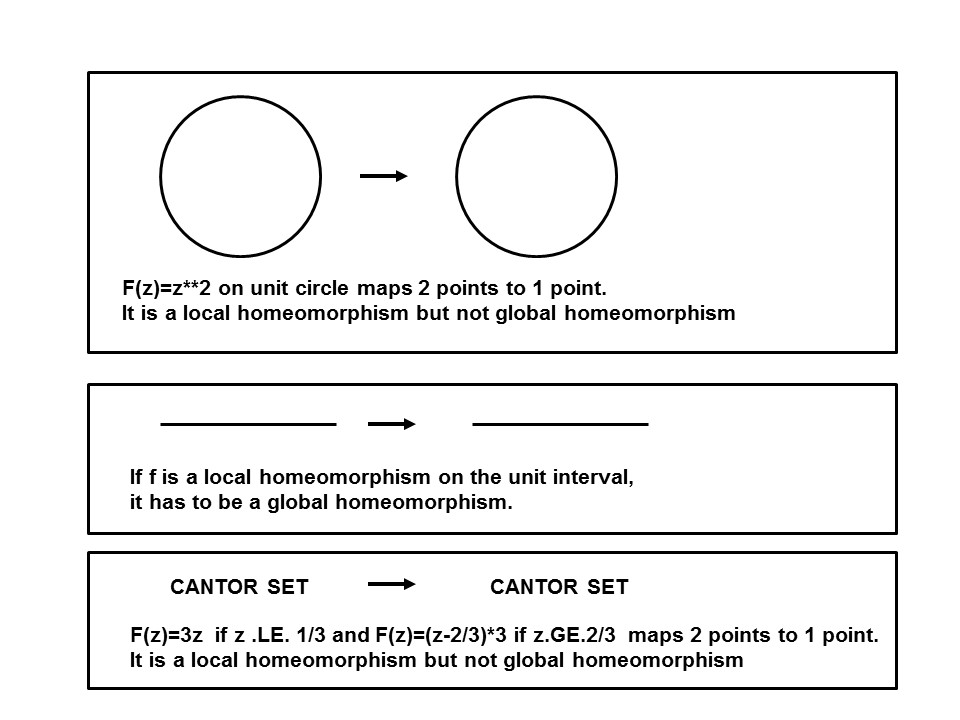}
\caption{Local homeomorphism (locally invertible function) is not always global homeomorphism (globally invertible).}
\label{fig1}
\end{figure}

Figure \ref{fig1} shows that local homeomorphism (locally invertible function) is not always global homeomorphism (globally invertible). Circle and Cantor set have ''holes'' which make the function local and not global homeomorphism.  The interval does not have holes so every local homeomorphism has to be a homeomorphism. 
 
This suggests that salt boundary can encounter difficulty with global inversion because it has hole. But we can locally invert it by breaking up the salt boundary by TOS, FOS, BOS which have no holes (TOS is top of salt, FOS is flank of salt and BOS is base of salt).

\begin{figure}
\centering
  \includegraphics[width=6in]{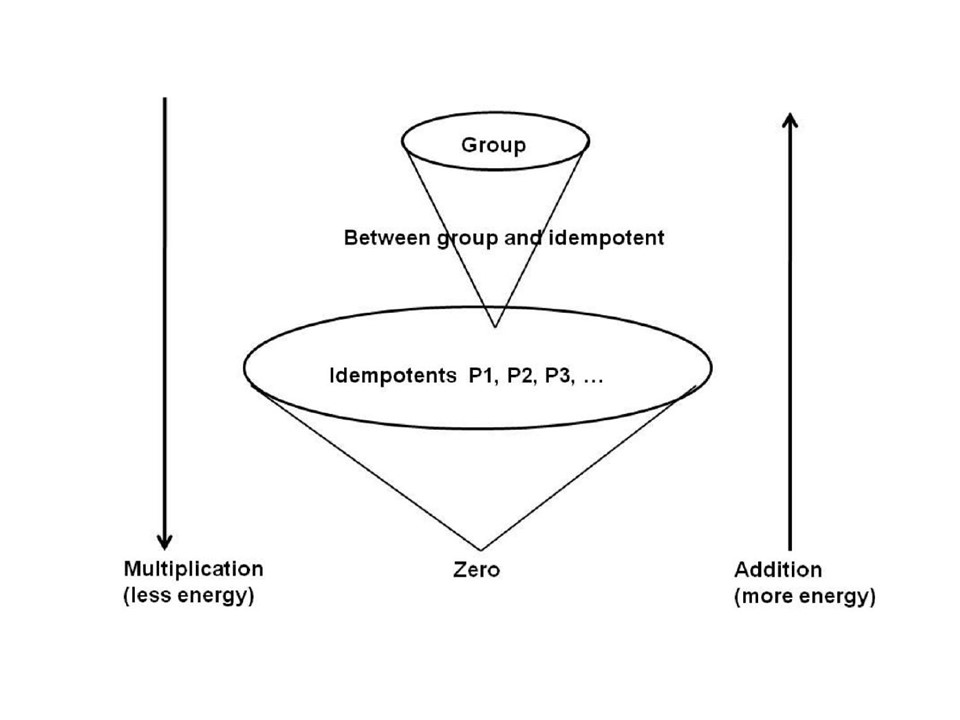}
\caption{Group theory (invertible system) is rare compared to semigroup theory (non-invertible or partially invertible system).}
\label{fig2}
\end{figure}

In ''Seismic Solvability Problems'', we showed a diagram, Figure \ref{fig2}, that group theory (invertible system) is rare compared to semigroup theory (non-invertible or partially invertible system).  Group theory is ''nice'' and has symmetry.  Semigroup theory is ''messy'' but describes field seismic data realistically.  Information in general experiences ''loss'' and ''attenuation'' due to earth properties which are not invertible.

\subsection*{QUALITATIVE METHOD}

Some of the difficulties with direct or indirect methods might be mitigated with qualitative methods.  Seismic inversion (inverse problem) could benefit from qualitative methods of interpretation, computational topology and simplified equation.

\subsection*{Qualitative method \# 1 (topology or geometry)}

Topological/geometric method (EULER):
Topology is simplification of seismic image into shapes in a qualitative way. This is the essence of Euler's view that a picture of bridges can be reduced to just lines and points.  Then compute the topological measure to define what is solvable and what is not.  This is a forerunner of modern computational topology and topological data analysis.

Surprisingly, seismic interpretation is following the same line of thought.  We take a complicated seismic image and draw surfaces and lines to simplify the picture and then converted them into maps to generate prospects to drill for oil and gas.

Seismic interpretation tries to simplify the seismic image into continuous part (horizons) and discontinuous part (faults or high contrast boundaries like gas pockets, salt, basalt, carbonate etc).

\subsection*{Application}

Computational topology attempts to measure discontinuities and holes (e.g. Betti numbers in homology). Topological measure is similar to geometric measure but it is not as rigid.  Cycle skipping is a common problem in quantitative inversion when single event splits into 2 events.  The 2 events have higher Betti number compared to single event so it is less desirable for minimizing Betti number.

\begin{figure}
\centering
  \includegraphics[width=5in]{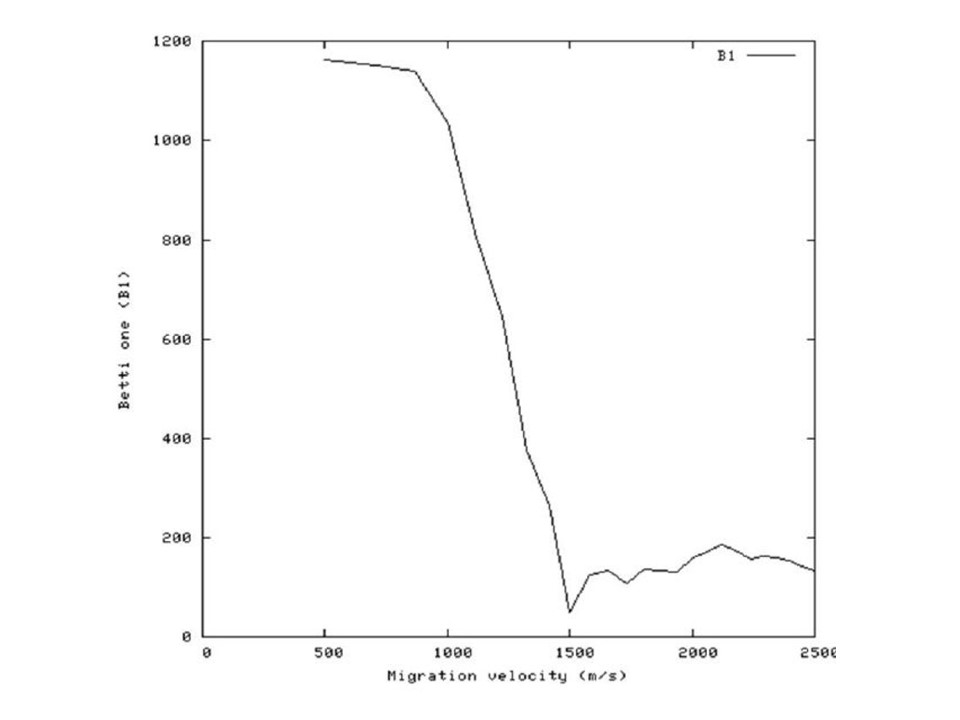}
\caption{An example of using Betti numbers to guide picking migration velocities.}
\label{fig3}
\end{figure}

In ''Seismic Solvability Problems'', we showed an example that Betti numbers could guide us into picking migration velocities in a qualitative way (Figure \ref{fig3}).   The lower Betti numbers have less complexity and tend to be better velocity picks. Computation topology or geometry is still a new research area.

In general, it is computationally intensive but it offers a different representation of real data.

\subsection*{Application (Velocity analysis with Betti number)}

\begin{figure}
\centering
  \includegraphics[width=5in]{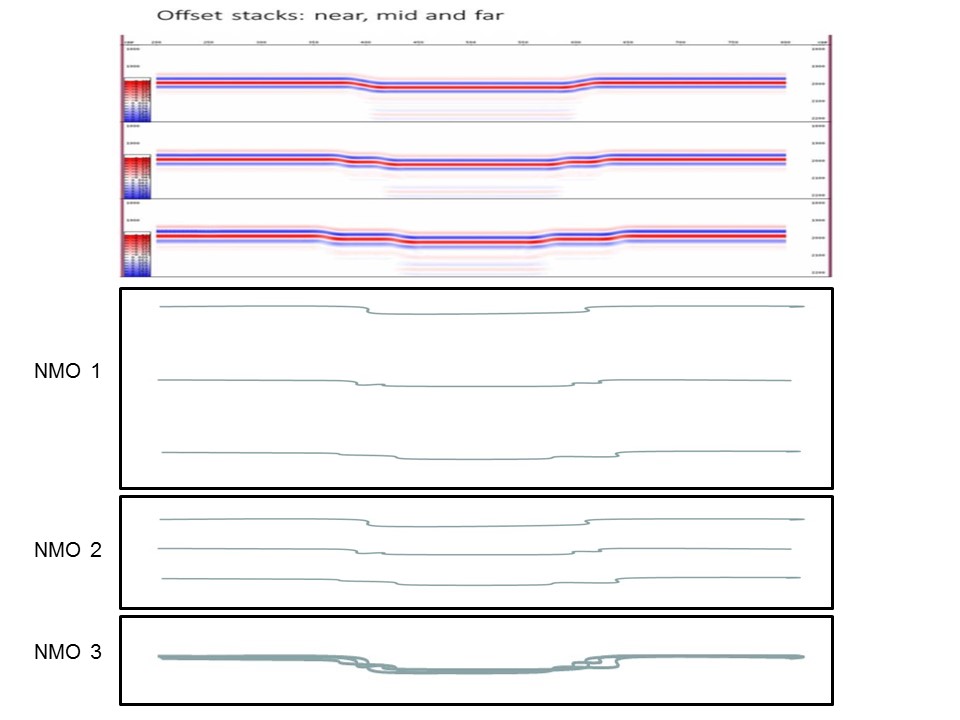}
\caption{Velocity analysis using common offset interpretation.}
\label{fig4}
\end{figure}

Another example of application of topology is velocity analysis using common offset interpretation (Figure \ref{fig4}).  If we interpret each common offset and the velocity gets closer to the final velocity,  the different horizons from different offsets will get closer.  For the difficult area,  the horizons intersect at a much higher Betti numbers.  This is a different measure from normal semblance which is algebraic.  The Betti number measures topological complexity.

\subsection*{Qualitative method \# 2 (Semigroup)}

Group theory is well known and Galois group theory is an elegant form of it.  However, for real field recorded data in seismic acquisition, it is normally noisy and has unrecoverable attenuation.  Hence it is difficult to describe in pure inverses as in a group.  

Semigroup is a more general concept adaptable to real data.  In functional analysis, semigroup of operators has been in use for a long time and we will not discuss in this paper. See Davies (2007).

The idea of semigroup is simple that the associative law holds:
$$ (a+b)+c = a+(b+c),$$
where + is an abstract symbol for an algebraic operation and not just addition.

Here are some examples of semigroups that are not groups:

If we take seismic traces $f$ and $g$,  then $f*g$ forms a semigroup where $*$ is convolution.

If we take the positive real numbers and use $+$ as normal addition,  it forms a semigroup.

If we take all subsets of the real number line,  then
$A+B = \{a+b:$ $a$ belongs to $A$ and $b$ belongs to $B\}$ 
forms a semigroup.
For example, $A=\{1,2\}$ and $B =\{4,6\}$ are two subsets,  then $A+B = \{5,7,6,8\} = \{5,6,7,8\}$.
Notice that $A$ has consecutive integers and $B$ does not.  Yet $A+B$ has consecutive integers.

\subsection*{Application (Cantor set)}

\begin{figure}
\centering
  \includegraphics[width=5in]{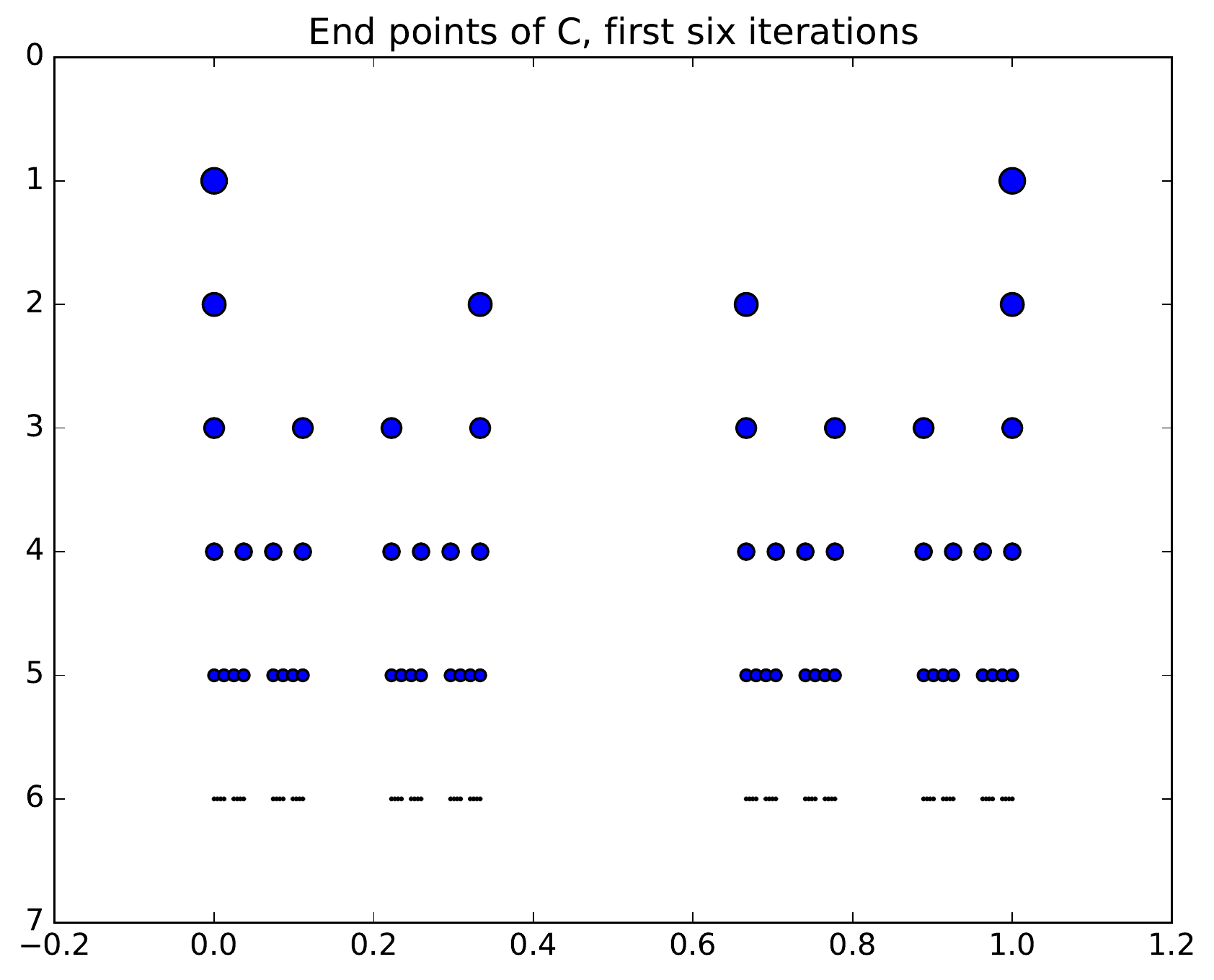}
\caption{End points in the Cantor set, in the  first 6 iterations}
\label{c1to6}
\end{figure}

\begin{figure}
\centering
  \includegraphics[width=5in]{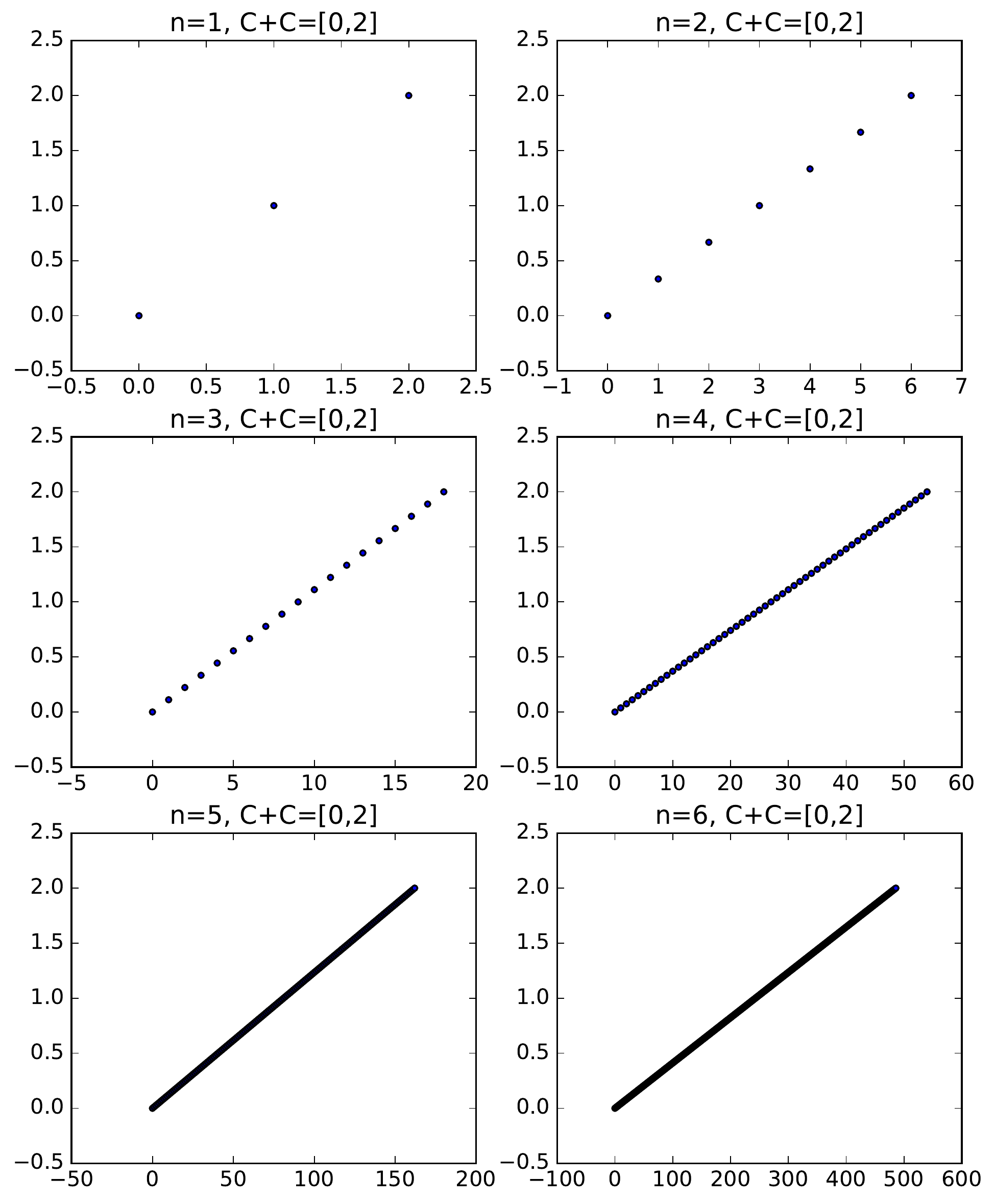}
\caption{A demonstration on the summing of two numbers from The Cantor set maps into a continum, i.e., $C + C = [0,2]$.}
\label{cpc}
\end{figure}

In Lau and Yin (2012),  we stated that $C+C = I$ where $C$ is the Cantor set and $I$ is a continuum.  Cantor is sparse since it does not contain any small continuous interval of numbers in $C$.  But the set sum of $C$ with itself is a continuous interval.  See Yamaguti et al (1997) Mathematics of fractals. Figures \ref{c1to6} and \ref{cpc} are a simple demonstration on this property.

\begin{figure}
\centering
  \includegraphics[width=5in]{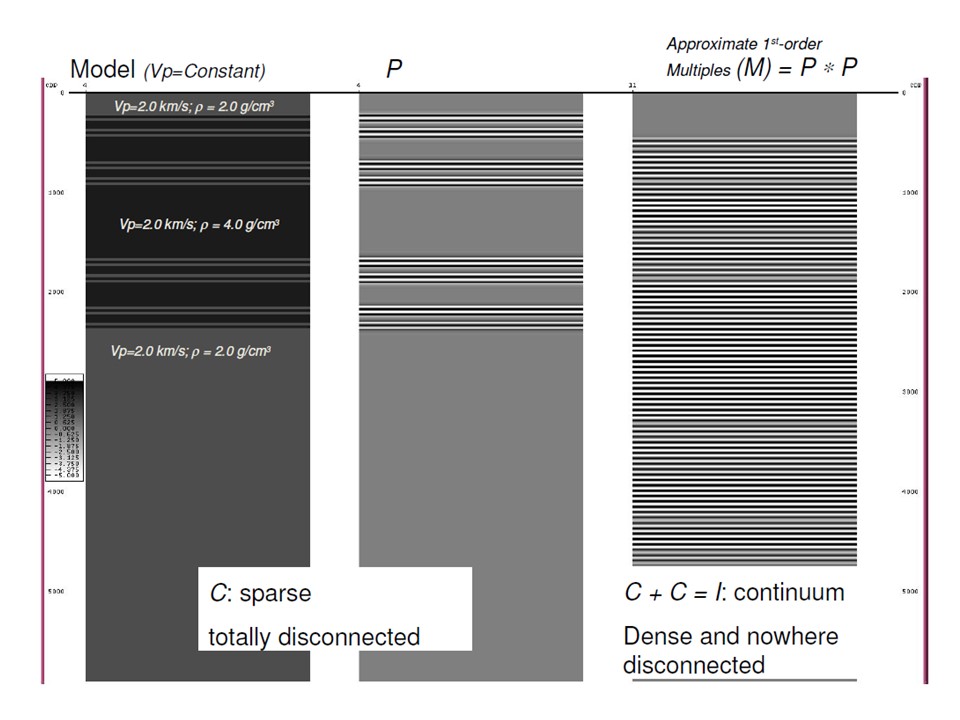}
\caption{If a primaries-only trace has support C,   then the surface multiples has support C+C.}
\label{fig5}
\end{figure}

So if a primaries-only trace has support $C$,   then the surface multiples has support $C+C$.  Figure \ref{fig5} illustrates that sparse support of the primaries could generate a continuum of multiples which is not invertible.  Indirectly, it demonstrates that multiples could only be removed for very sparse set.

\subsection*{Application (Diffusion semigroup)}

\begin{figure}
\centering
  \includegraphics[width=5in]{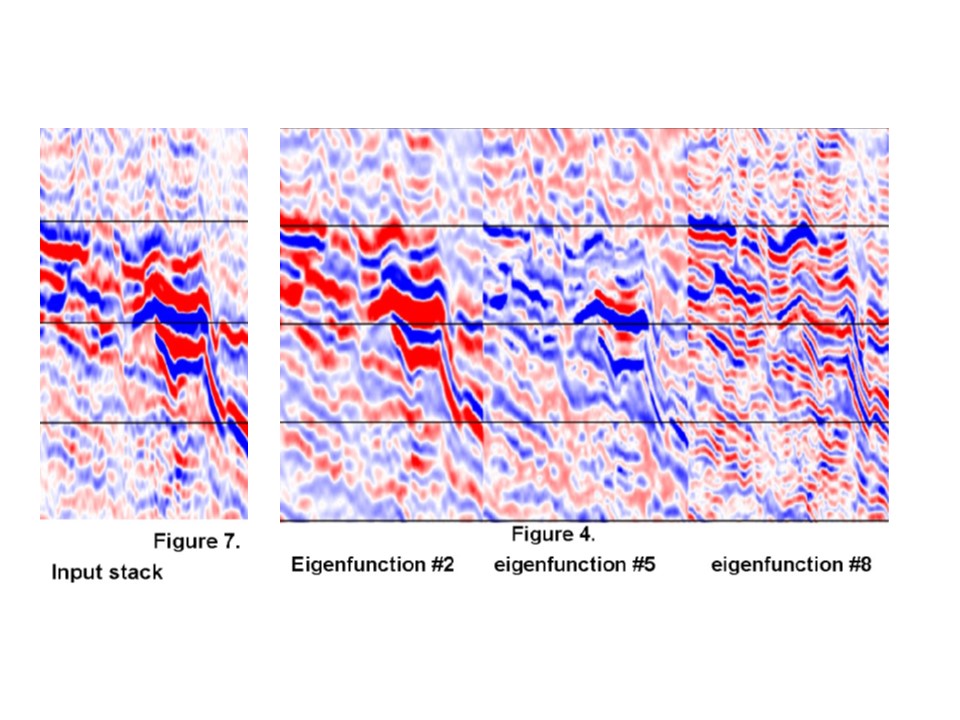}
\caption{The spectrum raised to an arbitrary power p is a semigroup.}
\label{fig6}
\end{figure}

Diffusion semigroup (sometimes called diffusion map) is a nonlinear decomposition of data  See Lau et al (2009). The generalized spectrum could be viewed as eigenfunctions (analog to Fourier spectrum for linear transform).  The spectrum raised to an arbitrary power p is a semigroup (Figure \ref{fig6}).  The power controls the diffusion of the data.  This is data diffusion and not physical diffusion.

\subsection*{Qualitative method \# 3 (Complex objects lead to lowering equation complexity)}

\begin{figure}
\centering
  \includegraphics[width=5in]{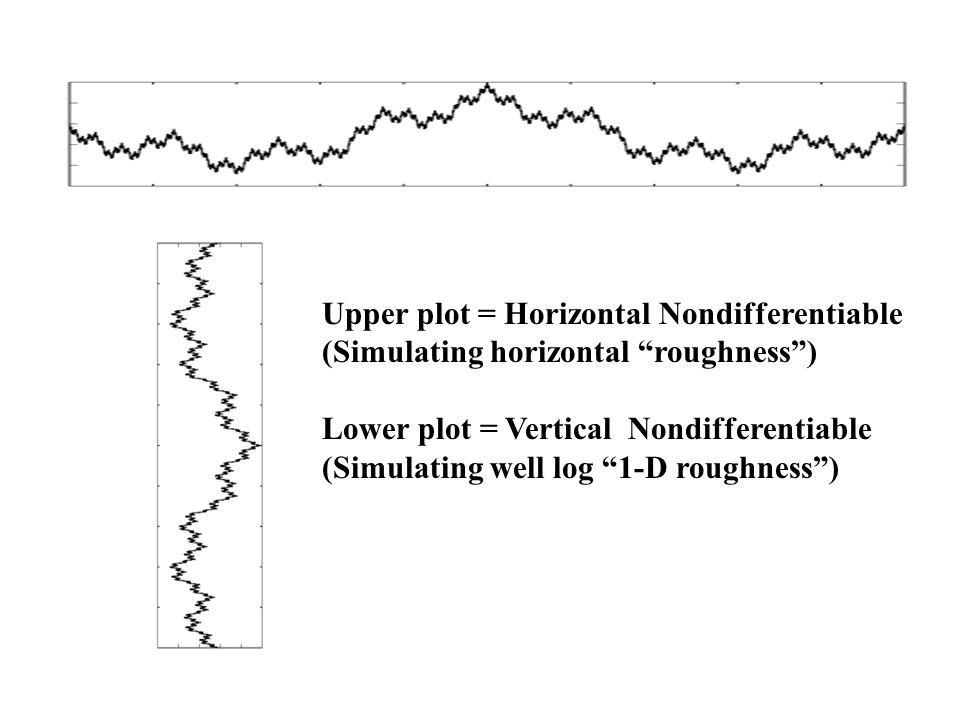}
\caption{Weierstrass function: a curve that is continuous but nowhere differentiable.}
\label{fig7}
\end{figure}

To motivate geometric complexity,  Weierstrass function is one of the ''pathological example'' of a curve that is continuous but nowhere differentiable.  It actually resembles ''fractal geology'' that had sharp erosions or cuts.  One of the characteristics of ''roughness'' is that it looks continuous but have many ''jumps'' on the curve (Figure \ref{fig7}).  Nowhere differentiable certainly satisfies the intuitive idea of many jumps.

Hutchinson's Theorem guarantees convergence of union of contraction sets.  The iterated function system (IFS) generates objects like Cantor set and Sierspinski carpet.  Even though there is similarity between geological feature and fractal set,  the fractal set does not fit real data exactly.  So we need some robust method of estimating roughness.

\subsection*{Application}

\begin{figure}
\centering
  \includegraphics[width=6.5in]{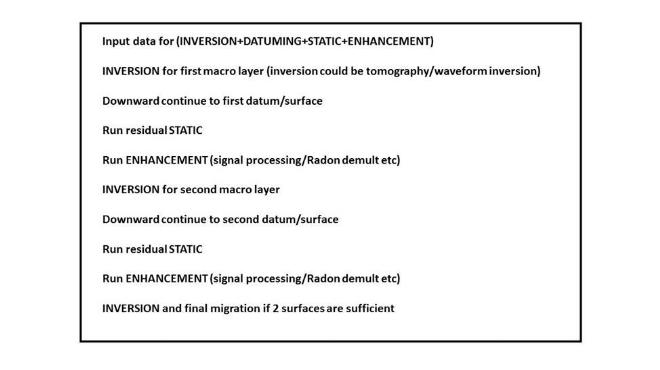}
\caption{A processing workflow of ''INVERSION + DATUMING + STATIC + ENHANCEMENT''.}
\label{fig9}
\end{figure}

In our paper ''Geometric theory of inversion and seismic imaging II: INVERSION + DATUMING + STATIC + ENHANCEMENT'', we offer the method of doing inversion for first macro-layer and then datum (wave equation or otherwise) to a smooth surface.  Then calculate static at new datum to reduce the equation from velocity and thickness to just traveltime.  Figure \ref{fig9} shows a suggested processing flow for this approach.

We have to reduce equation complexity from 2 terms of velocity and thickness to 1 term of traveltime. In some situations, we know that we cannot invert for thickness and velocity due to thin beds or rough surfaces beyond seismic resolution.  We showed that reducing the equation to time-shift (static) is beneficial. It is ''qualitative'' in the sense that we did not solve for ''physical parameters'' like thickness and velocity.  But it improves the imaging quality by reducing two terms to one.

For marine field data, there is also the problem of rough sea which does not obey the assumption of flat sea surface.  See Grion et al (2016).

\subsection*{CONCLUSION}

There is no one-size-fits-all in seismic processing.  

First: Examine the data with good diagnostics (QC plots) for quality control when we apply quantitative methods like direct/indirect inversion.  We have to be watchful of non-geologic results which could be meaningless or could lead to erroneous interpretation.

Second: Examine various prestack/poststack seismic images and discuss with end users (interpreters) the difficulty of the interpretation.  Apply qualitative methods of topology, semigroup and reducing equation complexity.   

\bibliographystyle{plain}
\bibliography{17qqsi_v4}  

\begin{thebibliography}{1}

\bibitem{}
Almasy, A., 2009, Inverse Problems in Classical and Quantum Physics, arXiv.

\bibitem{}
Davies, E. B., 2007, Linear Operators and their Spectra, Cambridge University Press.

\bibitem{}
 Grion, S., R. Telling and S. Holland, 2016, Rough sea estimation for phase-shift de-ghosting, 
 SEG International Exposition and 86th Annual Meeting.

\bibitem{}
Lau, A., 1975, The boundary of a semilattice on an n-cell,  Pacific Journal of Mathematics.

\bibitem{}
Lau, A., 1979, Certain local homeomorphisms of continua are homeomorphisms, II,  Bulletin Polish Acad. Science,  No.5, Vol XXVII. 

\bibitem{}
Lau, A., 1980, Plane Continuum and Transformation Groups, PAMS Proceedings of the American Society.

\bibitem{}
Lau, A., C. Yin, R. R. Coifman, A. Vassiliou, 2009,  Diffusion semigroups: A diffusion-map approach to nonlinear decomposition of seismic data without predetermined basis, SEG Annual Meeting.

\bibitem{}
Lau, A. and C. Yin, 2012, Seismic Solvability Problems, arXiv:1212.1350v1.

\bibitem{}
Lau, A. and C. Yin, 2017, Geometric theory of inversion and seismic imaging II: INVERSION + DATUMING + STATIC + ENHANCEMENT, viXra:1701.0303.

\bibitem{}
Lewis, W. and D. Vigh, 2016, 3D salt geometry inversion in full-waveform inversion using a level set method,  SEG Annual Meeting.

\bibitem{}
Virieux,J., R. Brossier, L. Métivier, S. Operto, A. Ribodetti,  2015 Preprint,  Direct and indirect inversions.

\bibitem{}
Yamaguti, M., M. Hata, and J. Kigami, 1997, Mathematics of fractals: Translations of
mathematical monographs: American Mathematical Society.	

\bibitem{}
Zhou, H., 2006, Multiscale deformable-layer tomography, Geophysics, vol. 71, No. 3,
May-June 2006, p. R11-R19.
\end{thebibliography}

\clearpage


\end{document}